\documentclass[prb,aps,twocolumn,amsmath,amssymb,floatfix,showpacs,
superscriptaddress]{revtex4}

\usepackage[dvips]{graphics}
\usepackage{color}
\definecolor{dred}{rgb}{0,0,0.6}

\begin{document}

\title{\textcolor{dred}{Unconventional low-field magnetic response of a 
diffusive ring with spin-orbit coupling}}

\author{Moumita Patra}

\affiliation{Physics and Applied Mathematics Unit, Indian Statistical
Institute, 203 Barrackpore Trunk Road, Kolkata-700 108, India}

\author{Santanu K. Maiti}

\email{santanu.maiti@isical.ac.in}

\affiliation{Physics and Applied Mathematics Unit, Indian Statistical
Institute, 203 Barrackpore Trunk Road, Kolkata-700 108, India}

\begin{abstract}

We report an unconventional behavior of electron transport in the limit
of zero magnetic flux in a one-dimensional disordered ring, be it completely
random or any correlated one, subjected to Rashba spin-orbit (SO) coupling.
It exhibits much higher circulating current compared to a fully perfect
ring for a wide range of SO coupling yielding larger electrical conductivity
which is clearly verified from our Drude weight analysis.

\end{abstract}


\maketitle

\section{Introduction}

The phenomenon of non-decaying circular current, the so-called persistent
current, in an isolated mesoscopic ring threaded by magnetic flux $\phi$
was first suggested by B\"{u}ttiker and his group~\cite{pc1} during early 
$80$'s. Following this proposal a considerable amount of theoretical work 
has been done~\cite{pc2,pc3,pc4,pc5,gefen,pc6,pc7,pc8,pc9,pc10,pc11,spl} 
towards this direction analyzing the effects of different 
factors like electron-electron (e-e) interaction, electron-phonon (e-ph) 
interaction, temperature, randomness, long-range hopping, spin-orbit (SO) 
couplings, etc. With these studies many significant features have been 
\begin{figure}[ht]
{\centering \resizebox*{4cm}{3.5cm}{\includegraphics{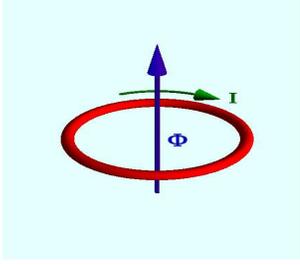}}\par}
\caption{(Color online). Schematic view of a $1$D ring carrying a current
$I$ in presence of magnetic flux $\phi$.}
\label{Model}
\end{figure}
explored those are consistent with experimental observations~\cite{ex1,ex2,
ex3,ex4,ex5,ex6} upto a 
certain level. But few debates still persist. For example, the appearance 
of discrepancy between current amplitudes in disordered rings obtained 
from experimental results and theoretical estimates. Measured currents 
are always higher than theoretical predictions. Much efforts have been
paid to resolve this issue and with some recent theoretical works 
possibilities of getting larger currents have been suggested which in 
some cases are very close to the experimental observations. The essential
factors those are responsible to enhance current in disordered rings 
are e-e interactions~\cite{pc3} and SO coupling~\cite{pc10}. Two types 
of SO couplings, namely Rashba~\cite{rash} and Dresselhaus~\cite{dres}, 
are usually encountered in solid state materials, out of which one 
(Rashba term) is controlled externally~\cite{meier}, whereas
the other is material dependent. Both these two coupling terms play 
identical role in enhancing current in a disordered ring as they are 
simply connected by a unitary transformation~\cite{pc10}, and due to 
this reason, in our present manuscript we consider only one of them.

It is well known that a disordered non-interacting (i.e., without any
e-e interaction) ring and free from any kind of SO coupling exhibits 
much smaller current~\cite{gefen} compared to a perfect ring. This is 
essentially due to electronic localization. But if we include SO 
interaction, a sufficient enhancement of current takes place~\cite{pc10} 
though it is always much less compared 
to a perfect ring. Now all these features have been studied for moderate
magnetic fluxes and, to the best of our knowledge, no one has investigated 
the response in the limit $\phi \rightarrow 0$. From our critical and deep
analysis we find an unconventional electronic behavior of a disordered ring 
in the limit of zero magnetic field subjected to SO coupling. Much higher 
current is obtained in a disordered ring than a completely perfect one for
a wide range of SO interaction, in contrast to the conventional literature
knowledge. This enhancement is also reflected in electrical conductivity
which we show by evaluating Drude weight.

The rest of our work is organized as follows. In Sec. II we present the
model and describe briefly the theoretical prescription. The results
are placed in Sec. III, and finally, we summarize in Sec. IV.

\section{Model and Theoretical Formulation}

Figure~\ref{Model} presents the schematic diagram of a one-dimensional
($1$D) ring threaded by a magnetic flux $\phi$ (measured in unit of 
$\phi_0=ch/e$). The tight-binding (TB) Hamiltonian of such a $N$-site
ring, subjected to Rashba SO coupling, reads as~\cite{pc12}
\begin{eqnarray}
\mbox{\boldmath$H$} &=& \sum_n \mbox{\boldmath $c$}_n^{\dag} 
\mbox{\boldmath $\epsilon$}_n \mbox{\boldmath $c$}_n  
+ \sum_n \left(e^{i\theta}\mbox{\boldmath $c$}_{n+1}^{\dag} 
\mbox{\boldmath $t$} \mbox{\boldmath $c$}_n + 
e^{-i\theta}\mbox{\boldmath $c$}_{n}^{\dag} 
\mbox{\boldmath $t$}^{\dag} \mbox{\boldmath $c$}_{n+1} \right) \nonumber \\
& & -\sum_n\alpha\left[\mbox{\boldmath $c$}_{n+1}^{\dag} 
\left(i\mbox{\boldmath $\sigma$}_x \cos\varphi_{n,n+1} + 
i\mbox{\boldmath $\sigma$}_y \sin\varphi_{n,n+1}\right)\right.\nonumber \\
& & \left. e^{i\theta} 
\mbox{\boldmath $c$}_n + h.c. \right] 
\label{equ2}
\end{eqnarray}
where $\alpha$ measures the strength of Rashba SO coupling and 
$\varphi_{n,n+1}=\left(\varphi_n + \varphi_{n+1}\right)/2$ with
$\varphi_n=2\pi(n-1)/N$ ($n$ is the site index). 
$\mbox{\boldmath $\sigma$}_x$,
$\mbox{\boldmath $\sigma$}_y$ and $\mbox{\boldmath $\sigma$}_z$ are
conventional Pauli spin matrices. Considering $c_{n\sigma}^{\dagger}$ 
($\sigma=\uparrow,\downarrow$) and $c_{n\sigma}$ as the creation and 
annihilation operators, respectively, we construct $\mbox{\boldmath $c$}_n$ 
and $\mbox{\boldmath $c$}_n^{\dagger}$ and they look like\\
$\mbox{\boldmath $c$}_n=\left(\begin{array}{c}
c_{n\uparrow} \\
c_{n\downarrow}\end{array}\right)$ and
$\mbox{\boldmath $c$}^{\dagger}_n=\left(\begin{array}{cc}
c_{n\uparrow}^{\dagger} & c_{n\downarrow}^{\dagger} 
\end{array}\right)$. 
$\mbox{\boldmath $t$}$ and $\mbox{\boldmath $\epsilon$}_n$ are both 
($2\times 2$) diagonal matrices where 
$\mbox{\boldmath $t$}_{11}=\mbox{\boldmath $t$}_{22}=t$ ($t$ being
the nearest-neighbor hopping element and during each hopping a phase factor
$\theta$ ($=2\pi\phi/N\phi_0$) is introduced), and $\epsilon_{n\uparrow}$
and $\epsilon_{n\downarrow}$ are two diagonal elements of 
$\mbox{\boldmath $\epsilon$}_n$, where $\epsilon_{n\sigma}$ corresponds to
the on-site energy. For a perfect ring $\epsilon_{n\sigma}$'s are constant
and we set them to zero without loss of any generality. On the other hand,
for a random disordered ring $\epsilon_{n\sigma}$'s 
($\epsilon_{n\uparrow}=\epsilon_{n\downarrow}$) are chosen randomly from 
a `Box' distribution function of width $W$ within the range $-W/2$ to 
$W/2$. As numerical results strongly depend on disordered configurations, 
we take the average over a large number of such distinct configurations.

To analyze the unconventional behavior of electron transport we need to
calculate current corresponding to each eigenstates along with net
current for a particular electron filling. The current carried by $m$-th
eigenstate is obtained from the following relation~\cite{pc12} 
\begin{eqnarray}
I_m & = & \frac{2\pi tie}{Nh}\sum\limits_{n}\left(
a_{n,\uparrow}^{m\ast}a_{n+1,\uparrow}^me^{-i\theta} + a_{n,\downarrow}^{m\ast}
a_{n+1,\downarrow}^me^{-i\theta} + h.c.\right) \nonumber \\
& & -\frac{2\pi\alpha e}{Nh}\sum\limits_{n}\left(
e^{-i\phi_{n,n+1}}a_{n,\uparrow}^{m\ast}a_{n+1,\downarrow}^me^{-i\theta}
\right.\nonumber \\
& & + \left.e^{i\phi_{n,n+1}}a_{n,\downarrow}^{m\ast}
a_{n+1,\uparrow}^me^{-i\theta} + h.c. \right).
\label{eq2}
\end{eqnarray}
Therefore, at absolute zero temperature net current carried by a ring 
containing $N_e$ electrons becomes $I=\sum_{m=1}^{N_e} I_m$.

Finally, the electrical conductivity is determined by calculating Drude
weight $D$ from the relation~\cite{pc13}
\begin{equation}
D=\frac{N}{4\pi^2} \left.\left( \frac{\partial^2 E_0(\phi)}{\partial \phi^2}
\right) \right|_{\phi \rightarrow 0}
\label{eq4}
\end{equation}
where $E_0(\phi)$ is the ground state energy. This parameter allows us
to predict the conducting ($D=$ finite) or insulating ($D \rightarrow 0$)
phase of any system.

Our main concern of this work is to study the interplay between $\alpha$,
$\phi$ and $W$ qualitatively, not quantitatively considering any specific
material, and therefore, throughout the numerical calculations we fix the
nearest-neighbor hopping integral $t$ at $1\,$eV and measure all other 
energies with respect to it. The current and Drude weight are calculated 
in units of $et/h$ and $e^2t/h^2$, respectively, where $e$ and $h$ are 
the fundamental constants.

\section{Results and Discussion}

\begin{figure}[ht]
{\centering \resizebox*{7cm}{11cm}{\includegraphics{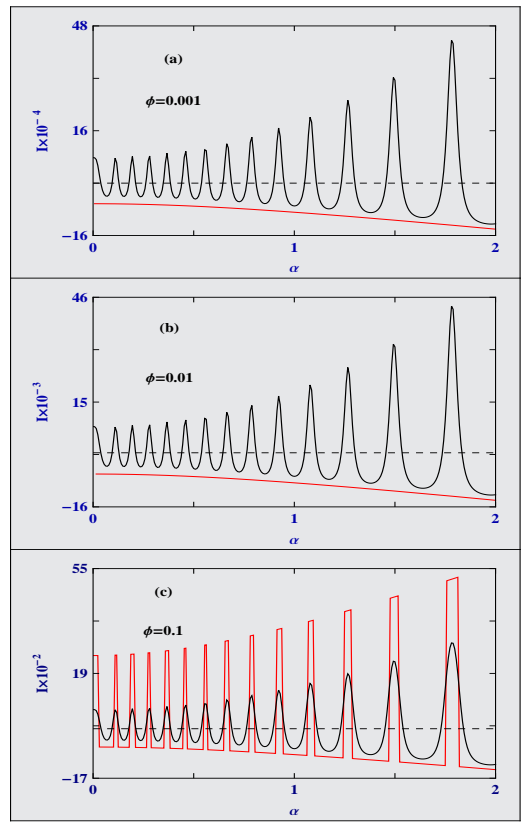}}\par}
\caption{(Color online). Dependence of current $I$ on $\alpha$ at three 
typical fluxes for a $80$-site ring considering $N_e=80$. The red and 
black curves correspond to the perfect and random disordered ($W=0.5$) 
rings, respectively. The dashed horizontal line represents
the line of zero current.}
\label{OD}
\end{figure}

\begin{figure}[ht]
{\centering \resizebox*{7cm}{4cm}{\includegraphics{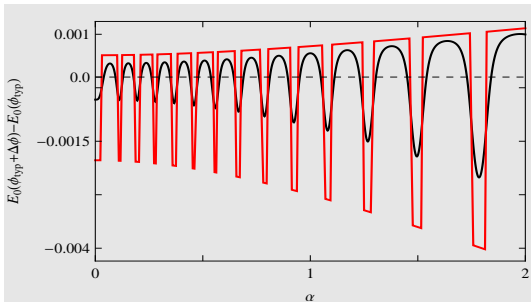}}\par}
\caption{(Color online). Variation of $\Delta E_0$ as a function of $\alpha$
at $\phi_{\mbox{\tiny typ}}=0.1$ for a $80$-site ring with $N_e=80$. The red 
and black curves correspond to the perfect and random disordered ($W=0.5$) 
rings, respectively. Here we set $\Delta \phi=0.125/16$.}
\label{deleng}
\end{figure}

Let us start with Fig.~\ref{OD} where the variation of current $I$ as 
a function of $\alpha$ is shown for a $80$-site half-filled ($N_e=80$)
ring. Three different cases, depending on $\phi$, are taken into account
and in each case we present the results of both ordered and random 
disordered rings. For a sufficiently low value of $\phi$ (viz, $\phi=0.001$)
current in the ordered ring increases monotonically, without showing any 
oscillation, with $\alpha$. While, anomalous oscillations with increasing 
peak heights are observed in the disordered ring (see Fig.~\ref{OD}(a)). 
Most notably we see that for a wide $\alpha$-window current in the 
disordered ring
becomes much higher compared to the perfect one, and the height of 
individual peak and its width get enhanced with increasing $\alpha$.
Exactly identical scenario, apart from greater current amplitude, also
persists when the flux $\phi$ moves to $0.01$ (see Fig.~\ref{OD}(b)).
This unconventional behavior i.e., appearance of much higher current in 
a disordered ring than a perfect one for several $\alpha$-windows 
disappears with increasing $\phi$ which is presented in Fig.~\ref{OD}(c).
Here, the current in the ordered ring is always higher for any 
$\alpha$-window than the disordered case and this conventional behavior
remains unchanged for any other higher values of flux $\phi$. Additionally,
we get oscillating nature of current in ordered case (not appeared in lower 
fluxes), where sharp jumps between positive and negative currents are 
observed, unlike the disordered ring which always exhibits a smooth 
variation over $\alpha$ (black line). This oscillating nature can be
implemented by analyzing the variation of $\Delta E_0$ as a function of
$\alpha$, where $\Delta E_0=E_0(\phi_{\mbox{\tiny typ}}+\Delta \phi)-
E_0(\phi_{\mbox{\tiny typ}})$ with $\Delta \phi \rightarrow 0_+$ (here
we choose $\Delta \phi=0.125/16$). For this
case $\phi_{\mbox{\tiny typ}}=0.1$. Using
this $\Delta E_0$, persistent current can be calculated through the 
relation $-\Delta E_0/\Delta \phi$ at $\phi=\phi_{\mbox{\tiny typ}}$
which is the conventional prescription~\cite{gefen} of determining 
persistent current
for a particular filling. Now from the variation of $\Delta E_0$ as a 
function of $\alpha$ (presented in Fig.~\ref{deleng}) we see that its sign 
alternates between positive and negative over a finite $\alpha$-window. 
Certainly, the factor ($\Delta E_0/\Delta \phi$) gives 
oscillating nature with sign reversal (red line of Fig.~\ref{OD}(c)) as a
function of $\alpha$ as $\Delta \phi$ is always positive. In addition it is 
important to note that the widths of the positive currents are much smaller
that the negative currents (though the scenario could be opposite i.e.,
lesser widths of negative currents than the positive ones for other fluxes).
This is solely associated with the variation of $\Delta E_0$-$\alpha$
characteristics. The window widths gradually decrease with lowering 
magnetic flux and eventually disappear for a sufficiently low values of
$\phi$ which results a continuous-like variation without providing any sharp
jump (red lines of Figs.~\ref{OD}(a) and (b)). On the other hand, 
disorderness makes a smooth variation of $\Delta E_0$ with respect to 
$\alpha$ (black curve of Fig.~\ref{deleng}) yielding a continuous-like 
pattern (without any sharp jump) of $I$-$\alpha$ characteristics (black 
lines of Fig.~\ref{OD}).

Regarding the oscillation in Fig.~\ref{OD} it is important 
to note that for all the three different fluxes current exhibits almost 
identical (though not exactly same) oscillating feature with $\alpha$. This
oscillation solely depends on the variation of $\Delta E_0$ with respect to
$\alpha$. For a particular ring size and fixed electron filling, the
$\Delta E_0$-$\alpha$ characteristics become almost similar for all fluxes,
since for any non-zero $\phi$ degeneracies are broken for a Rashba ring.
The oscillatory pattern (i.e., separation between successive peaks) could
be different for different filling factor associated with system size $N$. 
But, for a fixed $N$ and $N_e$ it is very hard to distinguish the separations
between the peaks for different fluxes, though they are actually different
which we confirm through our numerical calculations. In short, there is no 
proper periodicity of oscillation with $\alpha$. It depends on systems size 
$N$, filling factor $N_e$, etc., and further detailed studies should be
required to find if there is any specific periodicity.

\begin{figure}[ht]
{\centering \resizebox*{7cm}{7cm}{\includegraphics{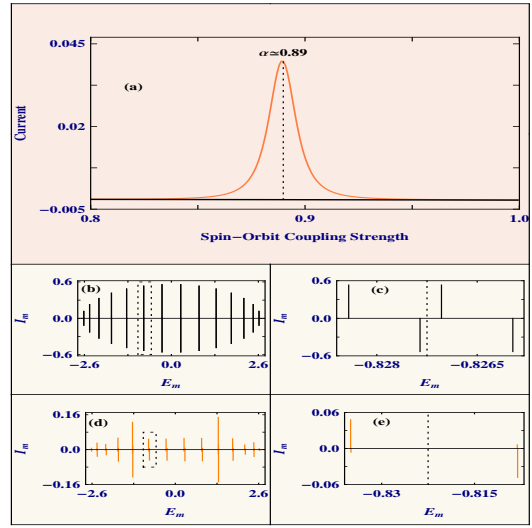}}\par}
\caption{(Color online). (a) Net current $I$ for a specific $\alpha$-window
considering a $30$-site ring with $N_e=24$, where the black and
orange curves correspond to ordered ($W=0$) and random disordered ($W=0.25$)
rings, respectively. (b) Current $I_m$ (vertical lines correspond to current
amplitudes) carried by $m$-th eigenstate having eigenenergy $E_m$ for the 
ordered $30$-site ring. (d) Same as (b) for the disordered ($W=0.25$) ring.
All these results are worked out for $\phi=0.001$, and the $I_m$-$E_m$
spectra are drawn for the critical $\alpha$ ($=0.89$) where $I$ has a peak.
The spectra (c) and (e) represent the zoomed version of the
dashed framed regions of (b) and (d), respectively, containing the state
currents of energy levels whose energies are close to Fermi energy $E_F$ 
associated with $N_e=24$. The
dashed vertical line is passing through the gap between the highest occupied 
energy level and the lowest unoccupied one. Below this dashed vertical line 
all states (i.e., $24$ states in each of the spectra (b) 
and (d)) are occupied which contribute to the net current, while the states 
above this line are empty. The current distributions of energy levels very 
close to Fermi energy, those essentially contribute to net current, can be
clearly seen from the spectra (c) and (e). The contributions of all other 
occupied states having lower energies mutually cancel with each other.}
\label{ODI}
\end{figure}

In order to explain the atypical behavior obtained in the limit
$\phi \rightarrow 0$ (i.e., higher current in disordered ring compared to 
the perfect one) let us focus on the spectra given in Fig.~\ref{ODI} 
where the results are computed for a $30$-site ring considering $\phi=0.001$. 
In Fig.~\ref{ODI}(a) the net currents for ordered (black line) and 
disordered (orange line) rings are superimposed for a specific 
$\alpha$-window. Now choose any $\alpha$, say $\alpha=0.89$, where the net
current in disordered ring is higher than the ordered one and try to
analyze the individual state currents for these two rings at this typical
$\alpha$ those are presented in Figs.~\ref{ODI}(b) and (d). In absence 
of impurities successive energy levels having a tiny difference in 
energy eigenvalues (as $\phi \rightarrow 0$, which is responsible to break
the degeneracy between $\pm k$ states) carry almost equal currents and in 
opposite directions (see Fig.~\ref{ODI}(b), where vertical lines correspond 
to the current amplitudes). On the other hand, when we add impurities 
currents carried by different states become more asymmetric with respect 
to each other, as clearly seen from Fig.~\ref{ODI}(d). Though all the energy 
levels upto Fermi energy $E_F$ associated with electron filling $N_e=24$ 
contribute to current, but the net contribution essentially comes from the 
energy levels having energies very close to $E_F$ as the contributions from 
\begin{figure}[ht]
{\centering \resizebox*{7cm}{7cm}{\includegraphics{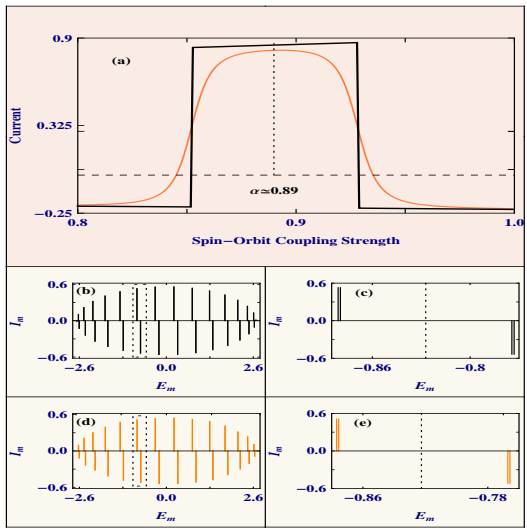}}\par}
\caption{(Color online). Same as Fig.~\ref{ODI} with $\phi=0.1$ (moderate
flux). Identical features are also observed for any other higher value of
$\phi$.}
\label{ODIhigh}
\end{figure}
all other energy levels mutually cancel with each other. For better viewing 
of the current distributions of energy levels close to Fermi energy, in 
Figs.~\ref{ODI}(c) and (e) we present the zoomed version of the framed
regions of Figs.~\ref{ODI}(b) and (d), respectively, where the dashed 
vertical line separates the occupied states from the unoccupied ones. 
Comparing the spectra given in Figs.~\ref{ODI}(c) and (e) we can easily
understand why net current in the disordered ring is higher than the 
perfect one though individual state currents are much less for the previous
ring. The asymmetric nature of current in the limit $\phi \rightarrow 0$ 
remains unchanged, yielding unconventional large current, for any kind of 
disordered ring (be it random or correlated) which we confirm through our 
detailed numerical calculations. This phenomenon can be argued physically 
as follows. There are three factors (SO coupling, magnetic flux and 
disorderness) which are responsible to control the current. In presence of
$\phi$ and $\alpha$, the current carried by individual states increases
{\em gradually} in the case of a pure ring as we move towards the energy
band centre. This is a well know phenomenon. While for the disordered ring
the gradual increment of current cannot be observed as mixture of high
conducting and low conducting states are there, which is even more 
transparent when $\phi$ is too small as it introduces very less current
in each state. Thus, much less conducting states exhibit vanishingly small 
currents compared to the other comparatively higher conducting states, which 
makes the $I_m$-$E_m$ spectrum asymmetric. This atypical nature is valid
for some specific $\alpha$-windows, whereas for other $\alpha$-regions we
get the conventional results associated with the interplay between $\alpha$,
$\phi$ and $W$.

From the above analysis atypical response of current in the low-field limit
($\phi \rightarrow 0$) is well understood. Now the question naturally comes
\begin{figure}[ht]
{\centering \resizebox*{8.5cm}{10cm}{\includegraphics{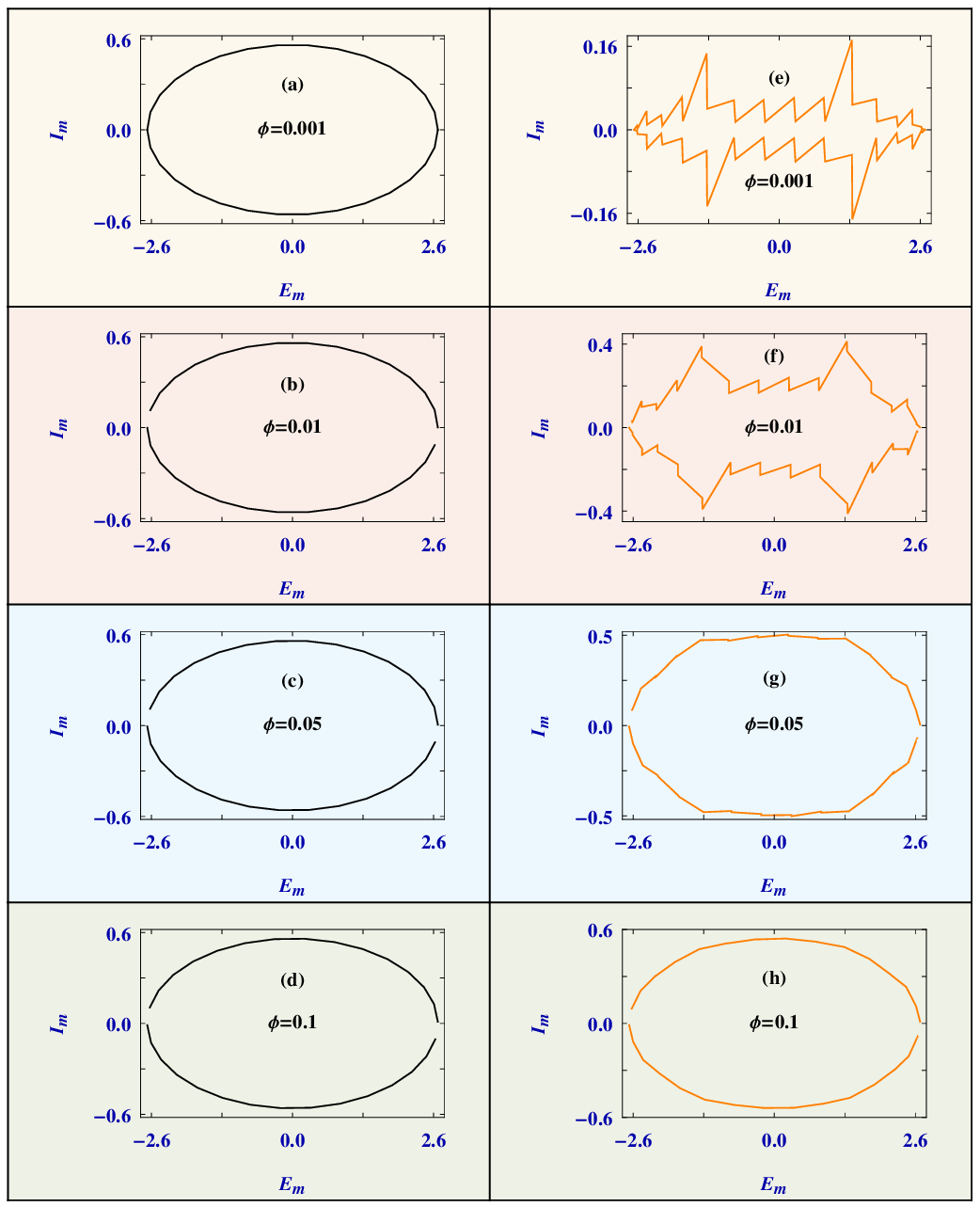}}\par}
\caption{(Color online). Envelope of the currents carried 
by individual energy levels for a $30$-site ring at some typical fluxes where 
the 1st and 2nd columns correspond to the ordered ($W=0$) and random 
disordered ($W=0.25$) rings, respectively. All positive currents carried
by different states are connected by a line to get one envelope along the
positive side, and similar prescription is given to get another envelope
along the negative side of individual spectrum.}
\label{env}
\end{figure}
why this behavior gets changed with increasing $\phi$, providing higher
current in a perfect ring compared to the disordered one for any $\alpha$.
This can be explained quite easily by studying currents carried by distinct
energy levels for both ordered and disordered cases. The results are shown
in Fig.~\ref{ODIhigh}. For moderate $\phi$ ($\phi \simeq 0.1$ or any higher 
value) all states carry much higher current compared to $\phi \rightarrow 0$ 
both for ordered and disordered cases (Figs.~\ref{ODIhigh}(b) and (d), 
respectively). The state currents for the perfect ring as usual symmetric
with respect to each other, but due to enhancement of currents as a result
of $\phi$ disordered ring also exhibits quite symmetric patterns in the
$I_m$-$E_m$ spectrum (Fig.~\ref{ODIhigh}(d)). Here it is interesting to note
that pairwise successive energy levels carry currents in opposite directions
which is more transparent from the zoomed version of the framed regions as
placed in Figs.~\ref{ODIhigh}(c) and (e), unlike the case of low-flux limit
where successive energy levels carry currents in opposite directions
(Figs.~\ref{ODI}(c) and (e)). Therefore, for a particular filling, a 
disordered ring exhibits smaller current than an ordered one, since for 
the later case individual state currents are always higher. 

From the spectra Figs.~\ref{ODI} and \ref{ODIhigh} we can
understand that a transition of $I_m$-$E_m$ spectrum from asymmetric
to symmetric nature takes place with the increment of magnetic flux $\phi$.
In order to visualize this transition with finer resolution in Fig.~\ref{env}
we present the envelop of the currents carried by individual energy levels
at some typical fluxes ranging from extremely low to a moderate one
where the first and second columns correspond to the perfect and disordered
rings, respectively. It is observed that in the limit of zero magnetic flux
positive and negative envelops are highly asymmetric for the disordered
ring and this asymmetric nature gradually decreases with increasing $\phi$. 
Eventually the asymmetry almost disappears for the moderate magnetic flux.
On the other hand, perfect ring always exhibits symmetric envelops as 
expected. From these results it can be emphasized that the interplay between 
$\alpha$, $\phi$ and $W$ is really very interesting and important too which 
has not been critically discussed before, to the best of our concern.

\begin{figure}[ht]
{\centering \resizebox*{7cm}{9cm}{\includegraphics{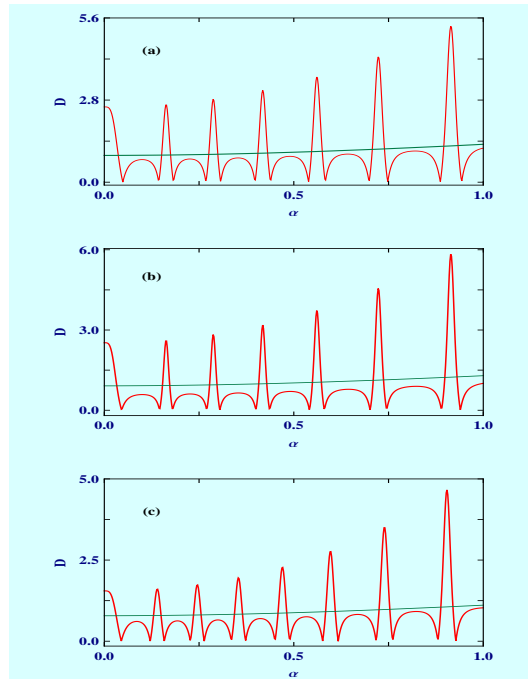}}\par}
\caption{(Color online). $D$-$\alpha$ characteristics (red line) of three 
different types of disordered rings. (a) Random disordered ($W=0.5$) ring 
with $N=55$, $N_e=54$; (b) $9$th generation ($N=55$) Fibonacci ring 
($\epsilon_A=-\epsilon_B=0.5$) with $N_e=54$ and (c) $7$th generation 
($N=64$) TM ring ($\epsilon_A=-\epsilon_B=0.25$) with $N_e=64$. In each 
case $D$-$\alpha$ curve (green line) for ordered ring 
($W=0$ or $\epsilon_A=\epsilon_B=0$) is superimposed.}
\label{DW}
\end{figure}

Finally, to substantiate more precisely the anomalous low-field response, 
in Fig.~\ref{DW} we present the variation of electrical conductivity 
(viz, Drude weight $D$) as a function of $\alpha$ both for ordered and 
disordered cases. Three different types of disordered rings, random,
Fibonacci and Thue-Morse (TM) are taken into account in order to establish 
the invariant nature of atypical response on disorderness in the limit 
$\phi \rightarrow 0$.
The results are presented in Figs.~\ref{DW}(a), (b) and (c), respectively,
and in each case $D$-$\alpha$ spectrum of ordered ring (green curve) is 
superimposed. Both these two correlated disordered rings (Fibonacci and 
Thue-Morse) are constructed using two primary lattices, (say) $A$ and $B$, 
following proper inflation rules. For the Fibonacci sequence~\cite{Fib} it 
is $A \rightarrow AB$ and $B \rightarrow A$, while for the other (TM) 
case~\cite{tm} it becomes $A \rightarrow AB$ and $B \rightarrow BA$. 
Depending on $A$-type or $B$-type atomic site $\epsilon_{n\sigma}$ can be 
simply written as $\epsilon_A$ or $\epsilon_B$, as sites are non-magnetic.
From the spectra (Fig.~\ref{DW}) it is observed that Drude weight exhibits 
pronounced oscillation in presence of disorder, similar to current 
oscillation in the limit $\phi \rightarrow 0$ (black lines of 
Figs.~\ref{OD}(a) and (b)), and this pattern does not change with 
disorderness i.e., whether it is correlated or random. On the other hand, 
for the ordered case a continuous variation with increasing magnitude 
of conductivity 
is obtained, obeying the earlier current analysis (red lines of 
Figs.~\ref{OD}(a) and (b)). Interestingly we see that for wide ranges 
of $\alpha$ electrical conductivity in disordered ring becomes much higher 
compared to the perfect one, and this strange behavior is fully consistent 
with our previous current analysis for $\phi \rightarrow 0$.

At the end, it is important to note that a disordered ring can provide 
metal-to-insulator transition at multiple values of $\alpha$ which is 
clearly visible from the oscillating nature of $D$-$\alpha$ curve (red
line of Fig.~\ref{DW}). Thus the present system can be utilized as a 
controlled switching device, since we can tune $\alpha$ externally, and
certainly it gives a high impact in the present era of nanotechnology.

The results studied here have been worked out for a Rashba ring. All these 
features remain exactly invariant if one takes a Dresselhaus ring instead 
of the Rashba one. In presence of Dresselhaus SO coupling the Hamiltonian 
of the ring looks like, 
\begin{eqnarray}
\mbox{\boldmath$\mathcal{H}$} &=& \sum_n \mbox{\boldmath $c$}_n^{\dag} 
\mbox{\boldmath $\epsilon$}_n \mbox{\boldmath $c$}_n  
+ \sum_n \left(e^{i\theta}\mbox{\boldmath $c$}_{n+1}^{\dag} 
\mbox{\boldmath $t$} \mbox{\boldmath $c$}_n + 
e^{-i\theta}\mbox{\boldmath $c$}_{n}^{\dag} 
\mbox{\boldmath $t$}^{\dag} \mbox{\boldmath $c$}_{n+1} \right) \nonumber \\
& & +\sum_n\beta\left[\mbox{\boldmath $c$}_{n+1}^{\dag} 
\left(i\mbox{\boldmath $\sigma$}_y \cos\varphi_{n,n+1} + 
i\mbox{\boldmath $\sigma$}_x \sin\varphi_{n,n+1}\right)\right.\nonumber \\
& & \left. e^{i\theta} 
\mbox{\boldmath $c$}_n + h.c. \right] 
\label{eq22}
\end{eqnarray}
where $\beta$ measures the strength of Dresselhaus SO coupling. Inspecting
the Hamiltonians for Rashba and Dresselhaus rings (Eq.~\ref{equ2}
and Eq.~\ref{eq22}) it is seen that these two Hamiltonians are connected via
a unitary transformation 
$U^{\dagger} \mbox{\boldmath$H$} U = \mbox{\boldmath$\mathcal{H}$}$,
where $U=(\mbox{\boldmath$\sigma$}_x + \mbox{\boldmath$\sigma$}_y)/\sqrt{2}$ 
is the unitary matrix. Therefore, any eigenstate $|\psi_m \rangle$ of
the Rashba ring can be written in terms of the eigenstate $|\psi_m^{\prime} 
\rangle$ of the Dresselhaus one where $|\psi_m \rangle = U |\psi_m^{\prime} 
\rangle$. This immediately gives the current for the Dresselhaus ring:
$I_m (\mbox{for}~ \mbox{\boldmath$\mathcal{H}$})= \langle \psi_m^{\prime}|
\mbox{\boldmath$I$}|\psi_m^{\prime} \rangle = \langle \psi_m|
U^{\dagger}\mbox{\boldmath$I$}U|\psi_m \rangle = \langle \psi_m|
\mbox{\boldmath$I$}|\psi_m \rangle = 
I_m (\mbox{for}~ \mbox{\boldmath$H$})$.
Therefore, the nature of the current carrying states for the Rashba ring 
becomes exactly identical to that of the Dresselhaus ring, and similar
argument is also true for other measurable quantities.

Though our analysis is purely theoretical, but all these features can
be verified experimentally since fabrication of small quantum ring with
few electrons (even less than ten electrons~\cite{keyser}) is possible 
with recent
technological advancements, and, the magnetic field required to produce 
such low fluxes (i.e, $\sim 0.001\phi_0$) is also within the experimental 
range. It is around $0.52\,$Tesla for a $100$-site normal metal 
ring with lattice spacing $a=1$A$^0$, which can definitely be achieved 
in realistic situation.

\section{Closing Remarks}

In the present work we have investigated an unconventional behavior of 
electron transport in a $1$D disordered mesoscopic ring subjected to
Rashba SO coupling. It provides a sufficiently large current compared to 
a fully perfect ring in the limit $\phi \rightarrow 0$ which has been 
analyzed by calculating individual state currents through second-quantized 
approach. The atypical response has been further confirmed by studying Drude
weight $D$. Like current, it also exhibits strange oscillations and for 
wide regions of $\alpha$ electrical conductivity in a disordered ring
becomes much higher than a perfect ring. Most notably we have seen that 
this atypical behavior is independent of the disorderness which we have 
verified by considering three different types of disordered rings. Finally,
a possibility of getting metal-to-insulator transition at multiple values
of $\alpha$ has been discussed that can be utilized for selective switching
action.

\section{Acknowledgment}

MP is thankful to University Grants Commission (UGC), India for
research fellowship.

\end{document}